\documentclass[12pt]{article}
\usepackage{latexsym,amssymb}

\begin{document}

\newcommand{\eqn}[1]{(\ref{#1})}
\newcommand{\eq}[1]{(\ref{#1})}

\newcommand{\bm}[1]{\mbox{\boldmath $#1$}}
\newcommand{\be}{\begin{equation}}
\newcommand{\ee}{\end{equation}}
\newcommand{\bea}{\begin{eqnarray}}
\newcommand{\eea}{\end{eqnarray}}
\newcommand{\nn}{\nonumber}
\newcommand{\ov}{\overline}
\newcommand{\ba}{\begin{eqnarray}}
\newcommand{\ea}{\end{eqnarray}}

\def\oddmarkC{\thepage}
\def\oddmarkB{}
\def\oddmarkD{}
\def\oddmarkE{}
\def\oddmarkF{}
\def\footmsgA{}
\def\sk{\vskip .4cm}
\def\ibar{{\bar \imath}}
\def\jbar{{\bar \jmath}}
\def\kbar{{\bar k}}
\def\lbar{{\bar \ell}}
\def\mbar{{\bar m}}
\def\Im{{\rm Im }}
\def\Re{{\rm Re }}
\def\IP{\relax{\rm I\kern-.18em P}}
\def\arccosh{{\rm arccosh ~}}
\def\Es{\bf E_{7(7)}}
\def\muun{\underline \mu}
\def\mun{\underline m}
\def\nuun{\underline \nu}
\def\nun{\underline n}
\def\buun{\underline \bullet}
\def\Eb{{\bf E}}
\def\bu{\bullet}
\def\we{\wedge}
\font\cmss=cmss10 \font\cmsss=cmss10 at 7pt
\def\twomat#1#2#3#4{\left(\matrix{#1 & #2 \cr #3 & #4}\right)}
\def\inbar{\vrule height1.5ex width.4pt depth0pt}
\def\IC{\relax\,\hbox{$\inbar\kern-.3em{\rm C}$}}
\def\IG{\relax\,\hbox{$\inbar\kern-.3em{\rm G}$}}
\def\IB{\relax{\rm I\kern-.18em B}}
\def\ID{\relax{\rm I\kern-.18em D}}
\def\IL{\relax{\rm I\kern-.18em L}}
\def\IF{\relax{\rm I\kern-.18em F}}
\def\IH{\relax{\rm I\kern-.18em H}}
\def\II{\relax{\rm I\kern-.17em I}}
\def\IN{\relax{\rm I\kern-.18em N}}
\def\IP{\relax{\rm I\kern-.18em P}}
\def\IQ{\relax\,\hbox{$\inbar\kern-.3em{\rm Q}$}}
\def\bfzero{\relax\,\hbox{$\inbar\kern-.3em{\rm 0}$}}
\def\IK{\relax{\rm I\kern-.18em K}}
\def\IG{\relax\,\hbox{$\inbar\kern-.3em{\rm G}$}}
 \font\cmss=cmss10 \font\cmsss=cmss10 at 7pt
\def\IR{\relax{\rm I\kern-.18em R}}
\def\ZZ{\relax\ifmmode\mathchoice
{\hbox{\cmss Z\kern-.4em Z}}{\hbox{\cmss Z\kern-.4em Z}}
{\lower.9pt\hbox{\cmsss Z\kern-.4em Z}} {\lower1.2pt\hbox{\cmsss
Z\kern-.4em Z}}\else{\cmss Z\kern-.4em Z}\fi}
\def\bfone{\relax{\rm 1\kern-.35em 1}}
\def\dop{{\rm d}\hskip -1pt}
\def\real{{\rm Re}\hskip 1pt}
\def\trace{{\rm Tr}\hskip 1pt}
\def\ii{{\rm i}}
\def\diag{{\rm diag}}
\def\sch#1#2{\{#1;#2\}}
\def\IU{\relax\,\hbox{$\inbar\kern-.3em{\rm U}$}}
\def\ls{{\Lambda\Sigma}}
\def\lg{{\Lambda\Gamma}}
\def\ld{{\Lambda\Delta}}
\def\sg{{\Sigma\Gamma}}
\def\sd{{\Sigma\Delta}}
\def\gd{{\Gamma\Delta}}
\def\sk{\vskip .4cm}
\def\noi{\noindent}
\def\ibar{{\bar\imath}}
\def\jbar{{\bar\jmath}}
\def\kbar{{\bar k}}
\def\al{\alpha}
\def\la{\lambda}
\def\be{\beta}
\def\ga{\gamma}
\def\Ga{\Gamma}
\def\de{\delta}
\def\epsi{\varepsilon}
\def\we{\wedge}
\def\part{\partial}
\def\bu{\bullet}
\def\ci{\circ}
\def\square{{\,\lower0.9pt\vbox{\hrule \hbox{\vrule height 0.2 cm
\hskip 0.2 cm \vrule height 0.2 cm}\hrule}\,}}
\def\muun{\underline \mu}
\def\mun{\underline m}
\def\nuun{\underline \nu}
\def\nun{\underline n}
\def\buun{\underline \bullet}
\def\Rb{{\bf R}}
\def\Eb{{\bf E}}
\def\gb{{\bf g}}
\def\dt{{\tilde d}}
\def\Dt{{\tilde D}}
\def\Dcal{{\cal D}}
\def\R#1#2{ R^{#1}_{~~~#2} }
\def\ome#1#2{\om^{#1}_{~#2}}
\def\Rf#1#2{ R^{\underline #1}_{~~~{\underline #2}} }
\def\Rfu#1#2{ R^{{\underline #1}{\underline #2}} }
\def\Rfd#1#2{ R_{{\underline #1}{\underline #2}} }
\def\Rfb#1#2{ {\bf R}^{\underline #1}_{~~~{\underline #2}} }
\def\omef#1#2{\om^{\underline #1}_{~{\underline #2}}}
\def\omefb#1#2{{ \omb}^{\underline #1}_{~{\underline #2}}}
\def\omefu#1#2{\om^{{\underline #1} {\underline #2}}}
\def\omefub#1#2{{\omb}^{{\underline #1} {\underline #2}}}
\def\Ef#1{E^{\underline #1}}
\def\Efb#1{{\bf E}^{\underline #1}}
\def\omb{\bf \mbox{\boldmath $\om$}}
\def\bfone{\relax{\rm 1\kern-.35em 1}}
\font\cmss=cmss10 \font\cmsss=cmss10 at 7pt
\def\a{\alpha} \def\b{\beta} \def\d{\delta}
\def\e{\epsilon}
\def\G{\Gamma} \def\l{\lambda}
\def\L{\Lambda} \def\s{\sigma}
\def\cV{\mathcal{V}}
\def\cA{{\cal A}} \def\cB{{\cal B}}
\def\cS{{\cal S}}
\def\cs{{\cal s}}
\def\cC{{\cal C}} \def\cD{{\cal D}}
\def\cF{{\cal F}} \def\cG{{\cal G}}
\def\cH{{\cal H}} \def\cI{{\cal I}}
\def\cJ{{\cal J}} \def\cK{{\cal K}}
\def\cL{{\cal L}} \def\cM{{\cal M}}
\def\cN{{\cal N}} \def\cO{{\cal O}}
\def\cP{{\cal P}} \def\cQ{{\cal Q}}
\def\cR{{\cal R}} \def\cV{{\cal V}}\def\cW{{\cal W}}
%
%
%
\def\crr{\crcr\noalign{\vskip {8.3333pt}}}
\def\tilde{\widetilde}
\def\bar{\overline}
\def\us#1{\underline{#1}}
\let\shat=\hat
\def\hat{\widehat}
\def\hyp{\vrule height 2.3pt width 2.5pt depth -1.5pt}
\def\Coeff#1#2{\frac{#1}{ #2}}
\def\Coe#1.#2.{\frac{#1}{ #2}}
\def\coeff#1#2{\relax{\textstyle {#1 \over #2}}\displaystyle}
\def\coe#1.#2.{\relax{\textstyle {#1 \over #2}}\displaystyle}
\def\half{{1 \over 2}}
\def\shalf{\relax{\textstyle \frac{1}{ 2}}\displaystyle}
\def\dag#1{#1\!\!\!/\,\,\,}
\def\to{\rightarrow}
\def\notin{\hbox{{$\in$}\kern-.51em\hbox{/}}}
\def\shdot{\!\cdot\!}
\def\ket#1{\,\big|\,#1\,\big>\,}
\def\bra#1{\,\big<\,#1\,\big|\,}
\def\equaltop#1{\mathrel{\mathop=^{#1}}}
\def\Trbel#1{\mathop{{\rm Tr}}_{#1}}
\def\inserteq#1{\noalign{\vskip-.2truecm\hbox{#1\hfil}
\vskip-.2cm}}
\def\attac#1{\Bigl\vert
{\phantom{X}\atop{{\rm\scriptstyle #1}}\phantom{X}}}
\def\exx#1{e^{{\displaystyle #1}}}
\def\del{\partial}
\def\delbar{\bar\partial}
\def\nex#1{$N\!=\!#1$}
\def\dex#1{$d\!=\!#1$}
\def\cex#1{$c\!=\!#1$}
\def\eg{{\it e.g.}} \def\ie{{\it i.e.}}
\def\IE{\relax{{\rm I\kern-.18em E}}}
\def\cE{{\cal E}}
\def\rt{{\cR^{(3)}}}
\def\IGam{\relax{{\rm I}\kern-.18em \Gamma}}
\def\IGa{\IA}
\def\ii{{\rm i}}
\def\inbar{\vrule height1.5ex width.4pt depth0pt}
\def\bfzero{\relax{\rm I\kern-.18em 0}}
\def\bfone{\relax{\rm 1\kern-.35em 1}}
\def\twomat#1#2#3#4{\left(\begin{array}{cc}
\end{array}
\right)}
\def\twovec#1#2{\left(\begin{array}{c}
{#1}\\ {#2}\\
\end{array}
\right)}
\def\o#1#2{{{#1}\over{#2}}}
\newcommand{\La}{{\Lambda}}
\newcommand{\Si}{{\Sigma}}
\newcommand{\im}{{\rm Im\ }}
\def\cA{{\cal A}} \def\cB{{\cal B}}
\def\cC{{\cal C}} \def\cD{{\cal D}}
\def\cF{{\cal F}} \def\cG{{\cal G}}
\def\cH{{\cal H}} \def\cI{{\cal I}}
\def\cJ{{\cal J}} \def\cK{{\cal K}}
\def\cL{{\cal L}} \def\cM{{\mathcal{M}}}
\def\cN{{\cal N}} \def\cO{{\cal O}}
\def\cP{{\cal P}} \def\cQ{{\cal Q}}
\def\cR{{\cal R}} \def\cW{{\cal W}}
\def\a{\alpha} \def\b{\beta} \def\d{\delta}
\def\e{\epsilon}
\def\G{\Gamma} \def\l{\lambda}
\def\L{\Lambda} \def\s{\sigma}
\def\T{T}

\begin{flushright}
CERN-PH-TH/2007-052\\
 UCLA/07/TEP/6
\end{flushright}
\vskip 1cm
  \begin{center}{\LARGE \bf  Black-Hole Attractors\\ \vskip 5mm in $N=1$ Supergravity}
\vskip 1.2cm
{L. Andrianopoli$^{1,2,3}$,  R. D'Auria$^2$, S. Ferrara$^{3,4}$ and M. Trigiante$^2
$}
\end{center}
 \vskip 3mm
\noindent
\begin{center}
{\small
$^1$ Centro E. Fermi, Compendio Viminale,
I-00184 Rome, Italy \\
$^2$
Dipartimento di Fisica,
  Politecnico di Torino, \\
  Corso Duca degli Abruzzi 24, I-10129
  Turin, Italy \\ and \\
  Istituto Nazionale di Fisica Nucleare (INFN)
  Sezione di Torino, Italy
  \\ \texttt{riccardo.dauria@polito.it}\\ \texttt{mario.trigiante@polito.it}\\
 $^3$
   CERN PH-TH Division, CH 1211 Geneva 23, Switzerland
  \\  \texttt{laura.andrianopoli@cern.ch}\\ \texttt{sergio.ferrara@cern.ch}
 \\
 $^4$ Istituto Nazionale di Fisica Nucleare (INFN)
  Laboratori Nazionali di Frascati\\ and \\
  Department of Physics and Astronomy, University of California,\\ Los Angeles, CA, USA}
\end{center}
\vfill

\vskip 1cm

\begin{abstract}
We study the attractor mechanism for $N=1$ supergravity coupled to vector and chiral multiplets and compute the attractor equations of these theories.
These equations may have solutions depending on the choice of the holomorphic symmetric matrix
$f_{\Lambda\Sigma}$ which appears in the kinetic lagrangian of the vector sector.
 Models with non trivial electric-magnetic duality group which have or have not attractor behavior are exhibited.
 For a particular class of models, based on an $N=1$ reduction of
 homogeneous special geometries, the attractor equations are
 related to the theory of pure spinors.
\end{abstract}

 \vfill\eject
\section{Introduction} \label{intro1}
The Reissner--Nordstrom charged black-hole  is a solution of the
Maxwell--Einstein system. This solution may have two horizons, one
horizon or no horizon  whenever $M^2 \gtreqless Q^2$, where $M$ is
the mass and $Q$ the charge of the black-hole.

In a supergravity context, such configuration can be either viewed as a particular solution of $N=2$ pure supergravity \cite{fvn} or of $N=1$ supergravity coupled to one vector multiplet \cite{fsv}.
Indeed, these theories have the same number of on-shell bosonic and fermionic degrees of freedom, but with a spin 3/2 gravitino exchanged with a photino. In the context of $N=2$ supergravity, the solutions with $M^2 \geq Q^2$ can be viewed as BPS or non-BPS \cite{gh}, while solutions with $M^2 < Q^2$ are forbidden (cosmic censorship) \cite{klo,bmo}.

In the $N=1$ theory, the bosonic solutions are the same, so $M^2 <
Q^2$ is still forbidden in spite of the fact that  no
supersymmetric black-holes exist in this case. For $M^2=Q^2$ the
horizon geometry is Bertotti--Robinson, with a $AdS_2 \times S_2$
metric \cite{bero}.

Recent investigation (for recent reviews, see for instance
\cite{moh,dew,pio,veneziano,bfm}) have in fact shown that extremal
black-holes with attractor behavior also exist without saturating
the BPS bound \cite{ortin,fgk,sen,gijt,k1,k2}. Many examples in
$N=8$ supergravity \cite{fk06} as well as in generic $N=2$
theories have been given \cite{tt}, so that such configurations
may be studied also in theories which do not have BPS black-hole
configurations \cite{freed,cd07}.

The aim of this investigation is to consider particular theories
of $N=1$ supergravity coupled to matter multiplets, which may have
extremal black-hole solutions with attractor behavior
\cite{fks,strom,fk}. We will extend the analysis to $N=1$ theories
with scalar fields, where extremal black-holes are connected to
attractor points for the scalars.

Now we have an unspecified number $n_V$ of vector multiplets ($\Lambda =1,\cdots n_V$) and $n_S \geq 1$ of chiral multiplets. The electric-magnetic duality properties of these lagrangians have been studied in \cite{cdfv}, following the general analysis given in \cite{gazu}.
In the general case of a theory coupled to chiral and vector multiplets, to have a consistent solution exhibiting attractor behavior, the crucial element  is encoded in a complex symmetric matrix, $f_{\Lambda\Sigma}$, with $\Im  f < 0$, which is related to the kinetic term of the gauge fields \cite{cfgv}
\begin{equation}
 \sqrt {g}\Im f_{\Lambda\Sigma} F^\Lambda_{\mu\nu} F^{\Sigma |\mu\nu} + \frac 1{2} \Re f_{\Lambda\Sigma} F^\Lambda_{\mu\nu}  F^{\Sigma}_{\rho
\sigma}\epsilon^{\mu\nu\rho\sigma}.
\end{equation}
The matrix $f_{\Lambda\Sigma}$ must satisfy some particular properties, in particular it has to be a holomorphic function of the scalar fields $\partial_\ibar f_\ls =0$.

In terms of $f$ the black-hole potential reads \cite{fgk}
\begin{equation}
V =-\frac 12 (q_\Lambda - f_\ls p^\Sigma) (\Im
f^{-1})^{\Lambda\Gamma} (q_\Gamma -\bar f_{\Gamma\Delta}
p^\Delta)=-\frac 12 Q^T \mathcal{M}(f) Q
\label{bhpot}\end{equation} with $Q=(p^\Lambda , q_\Lambda )$ the
(constant) charge vector and $\mathcal{M}$ the symmetric,
symplectic, negative defined matrix ($\cM^T=\cM$, $\cM \cdot
\Omega \cdot \cM = \Omega$, where $\Omega$ is the
$Sp(2n_V,\mathbb{R})$ invariant metric $\pmatrix{0&-\bfone\cr
\bfone & 0}$) given by
\begin{equation}
\mathcal{M} =\pmatrix{\Im f + \Re f \Im f^{-1} \Re f & -\Re f \Im f^{-1}\cr
-\Im f^{-1} \Re f & \Im f^{-1}}.
\end{equation}
To have large extremal black-hole solutions we require that the black-hole potential has an extremum $\partial_i V =0$ at $V|_{\mbox{extr}}\neq 0$, with Hessian matrix
$\partial\partial V$ positive definite.
The black-hole entropy is then given by \cite{fgk}:
\begin{equation}
S_{BH}(p,q)=\pi V|_{\partial_i V =0}.\label{entropy}
\end{equation}

In $N=1$ theories the vector kinetic matrix $f_\ls$ is not fixed by supersymmetry and it can in principle be a rather arbitrary holomorphic function of the chiral multiplets.
However, in theories which originate from higher dimensions, such as the ones coming from superstring compactifications, the matrix $f_\ls$ may have a restricted form due to the symmetries of the theory.

For instance, in section 2 we will consider particular $N=2$ models whose bosonic sector coincides (without truncations) with $N=1$ models and which exhibit non-trivial attractor behavior. Examples of $N=1$ models which have no higher $N$ analogue can be obtained for
 Grassmannian  manifolds $U(n,n)/[U(n) \times U(n)]$, following the results of \cite{az}.

As another example, in heterotic string compactifications on
Calabi--Yau threefolds the tree-level form of $f_\ls$  is just
$f_\ls = S \delta_\ls$ where $S$ is the chiral dilaton-axion
multiplet \cite{wit}. This is the first example we will encounter
in section 3. For Calabi--Yau orientifolds (in type IIA) this
matrix is linear in the chiral fields, and a class of examples which
share similar properties are the models coming from
(orientifolded) homogeneous special geometries. Their general
features will also be discussed in section 3. The $N=1$ theories
obtained from truncation of homogeneous special geometries exhibit
the particular feature that the chiral multiplets sector is
described by a non linear $\sigma$-model of the type
$SO(2,n)/[SO(2)\times SO(n)]$ and the vector multiplets are in
several copies of the spinor representation of $spin(1,n-1)$
which, combining electric and magnetic field-strengths extends to
the electric-magnetic duality group $spin(2,n)$. These theories
generally show attractor behavior and their critical points have
the nice geometrical interpretation that a certain
moduli-dependent spinor, constructed out of the electric and
magnetic charges,  becomes a pure spinor\footnote{We call here,
with  an abuse of language, a pure spinor a spinor $\psi$ for
which $\psi \gamma_\mu \psi =0$ \cite{pure}.}.  At the end of the
section, the Hessian matrix of some $N=1$ models at the critical
points is determined, showing the attractor nature of the
solution. The paper ends in section \ref{conclusions} with some
concluding remarks.


\section{Embedding  Maxwell--Einstein theory in $N=1$ supergravity and attractors}\label{embedding}
We are going to discuss in this section the conditions to have
extremal black-hole attractor solutions (for large black-holes) in
theories with $N=1$ supergravity.

The crucial condition on the scalar sector is the request of an holomorphic matrix $f_\ls$. The attractor equation $\partial_i V=0$, for arbitrary matrix $f_{\Lambda\Sigma}$ (with $\Im f_\ls <0$) satisfying $\partial_\ibar f_\ls =0$ may be written as
\begin{equation}
\partial_i V =0 = \bar\cV^\Lambda \partial_i f_\ls \bar \cV^\Sigma \label{min}
\end{equation}
where:
\begin{equation}
\cV_\Lambda \equiv (q_\Lambda - f_\ls p^\Sigma)\,, \qquad \partial_\ibar \cV_\Lambda=0\,; \qquad \cV^\Lambda = (\Im f^{-1})^\ls \cV_\Sigma ,
\end{equation}
and the inverse formula holds:
\begin{eqnarray}
p^\Lambda = -\Im \cV^\Lambda\,,\qquad
q_\Lambda = \Re \cV_\Lambda - \Re f_{\ls} \Im \cV^\Sigma \label{inverseformula}
\end{eqnarray}

Indeed,
\begin{equation}
 V =-\frac 12\bar \cV_\Lambda  (\Im f^{-1})^\ls \cV_\Sigma =
 -\frac 12 \bar \cV^\Lambda  \Im f_\ls  \cV^\Sigma =-\frac 12  \bar\cV_\Lambda
 \cV^\Lambda\label{potV}
\end{equation}
and
\begin{equation}
\partial_i  \cV^\Lambda =-\frac 1{2\ii} (\Im f^{-1})^\ls \partial_i f_\sg \bar \cV^\Gamma
\label{dvl}
\end{equation}
so that
\begin{equation}
\partial_i  V =-\frac 1{2} \bar\cV_\Lambda  \partial_i  \cV^\Lambda =\frac 1{4\ii} \bar \cV^\Lambda \partial_i f_\ls \bar\cV^\Sigma
\end{equation}
We consider large black-holes solutions, then we require that at
the attractor point eq. \eq{min} be satisfied with $\bar
\cV^\Lambda \neq 0$. The interpretation of the black-hole
potential in eq. (\ref{potV}) is that $\mathcal{V}^\Lambda$ is the
shift which appears in the photino supersymmetry variation
$\delta_\epsilon \lambda^\Lambda$ in the presence of the charged
black-hole background. A bilinear photino-gravitino term
\cite{cfgv} in the geodesic action \cite{fgk} with field strengths
$\mathcal{V}^\Lambda$ shows that supersymmetry is spontaneously
broken \cite{fk,bfm2}.

Examples of non-linear $\sigma$-models for  chiral multiplets which are compatible with a non-trivial electric-magnetic duality of the Maxwell fields \cite{gazu} (that is with a scalar-dependent holomorphic matrix $f_\ls$) are
\begin{enumerate}
\item{ $Sp(2n,\mathbb{R})/U(n)$ coupled to $n$ vector multiplets, with duality group $ Sp(2n,\mathbb{R})$}
\item{$U(1,n)/U(n)$  coupled to $n+1$ vector multiplets, with duality group $U(1,n) \subset Sp(2n+2, \mathbb{R})$}
\item{ $SU(1,1)/U(1)$ coupled to $n$ vector multiplets, with duality group $SL(2,\mathbb{R})) \times SO(n) \subset Sp(2n,\mathbb{R})$}
\item{ $ SO(2,n)/SO(2)\times SO(n)$ coupled to $r$ vector multiplets in the spinor representation of $SO(1,n-1) \subset SO(2,n)$, with duality group $spin(2,n) \subset Sp(2r,\mathbb{R})$, where $r$ is the dimension of the spinor representation of $SO(1,n-1)$}
\item{$U(n,n)/U(n)\times U(n)$ coupled to $2n$ vector multiplets}
\end{enumerate}
As we will see in the next sections, examples 2,3,4,5 exhibit in
general attractor behavior, while example 1 does not. This can be
easily understood because in the $Sp(2n,\mathbb{R})/U(n)$ case the
scalar fields $x_\ls = x_{\Sigma\Lambda}$ belong to the symmetric
representation of $U(n)$, and we have, for the kinetic matrix of
the vector
\begin{equation}
f_\ls =x_\ls .
\end{equation}
Then, from \eq{min} we find that the attractor equation for this model is
\begin{equation}
\partial_\ls V =0 \; \Rightarrow \; \bar\cV_\Lambda \bar\cV_\Sigma =0
\end{equation}
whose only solution is $\cV_\Lambda =0$, which implies
$V|_{\mbox{extr}}=0$. This solution may correspond to a small
black-hole, while attroctor solutions for large black-holes cannot
be found for this model.


\subsection{$N=1$ theories with special geometry}

An attractor behavior is guaranteed  in theories where the kinetic matrix $f_{\Lambda\Sigma}$ is defined  in a special-K\"ahler geometry.
First of all, to have an $N=1$ theory with special geometry for the scalar sector, it is necessary that the number of  Wess--Zumino multiplets and of vector multiplets be related. In particular, if the number of chiral multiplets is $n_S =n$,  the number of vector multiplets has to be $n_V=n+1$.
Then, the following identity has to hold \cite{cdf,fk}:
\begin{equation}
V=-\frac 12 Q^T \mathcal{M} Q = |D_i Z|^2 + |Z|^2,
\end{equation}
in terms of a covariantly holomorphic superpotential ($N=2$ central charge)
\begin{equation}
Z(z)=e^{\frac{\mathcal K}2 } (X^\Lambda q_\Lambda - F_\Lambda p^\Lambda )\,,\qquad X^\Lambda =(1,z^i)\,, \quad i=1,\cdots n.
\end{equation}
Using the relations of special geometry the attractor condition is in this case
\begin{equation}
\partial_i V =0= 2 \bar Z D_i Z +\ii C_{ijk} \bar Z^{j} \bar Z^{k} \,, \qquad (\bar Z^j \equiv g^{j\ibar}D_\ibar \bar Z).\label{special}
\end{equation}

However, for the theory to be $N=1$ supersymmetric the matrix $f_{\Lambda \Sigma}$ must be holomorphic (for a given choice of coordinates).
But for general special-K\"ahler geometries, the kinetic matrix for the vectors, $\cN_\ls$,
which is related to the covariantly holomorphic symplectic section $U^M{}_0= (L^\Lambda , M_\Lambda ) = e^{\frac{\mathcal K}2 } (X^\Lambda ,F_\Lambda )$ and to its covariant derivative $\bar U^M{}_i=D_i U^M{}_0 \equiv (f^\Lambda_i , h_{\Lambda i})$  via
\begin{eqnarray}
\left\{\matrix{M_\Lambda &=& \cN_\ls L^\Sigma \hfill\cr
h_{\Lambda i} &=& \bar\cN_\ls f^\Sigma_i\hfill}\right.,\label{spegeo}
\end{eqnarray}
is in general neither holomorphic nor antiholomorphic.
We find indeed, from \eq{spegeo}
\begin{eqnarray}
\left\{\matrix{(\partial_i \cN_\ls)L^\Sigma &=& -(\cN -\bar\cN)_\ls f^\Sigma_i \hfill \cr
(\partial_i \cN_\ls)\bar f^\Sigma_\jbar &=&0 \hfill}\right.\label{antiholn}
\end{eqnarray}
and
\begin{eqnarray}
\left\{\matrix{(\partial_\ibar \cN_\ls)L^\Sigma &=&0\hfill\cr
(\partial_\ibar \cN_\ls)\bar f^\Sigma_\jbar &=& \ii C_{\ibar\jbar\kbar} g^{\kbar\ell} (\cN -\bar\cN)_\ls f^\Sigma_\ell\hfill}\right.\label{holn}
\end{eqnarray}
From \eq{holn} we find that, for the case $n_V =n_S +1$, the only way to have a holomorphic kinetic matrix is to make the identification $\cN_\ls =f_\ls$ and ask $C_{ijk}=0$, in which case we have $\partial_\ibar f =0$.
The bosonic sector of the theory found in this way is then an $N=1$ model which is identical to the one of an $N=2$ model \footnote{This is the so-called minimal coupling of $n$ vector multiplets to $N=2$ supergravity \cite{luc}.}.
The only way to satisfy the above properties is to consider as non-linear $\sigma$-model spanned by the scalar sector the series
$\frac{U(1,n)}{U(1)\times U(n)}$.
For this series of special-K\"ahler models indeed $C_{ijk}=0$,  and the kinetic matrix $\cN_\ls$ is holomorphic. In the basis with prepotential $F(X) =-\frac\ii 2 \eta_\ls X^\Lambda X^\Sigma$ ($\eta_\ls = (1,-1,\cdots,-1)$) we have
\begin{equation}
\cN_\ls = \ii\left(\eta_\ls -2 \frac{X_\Lambda X_\Sigma}{X^2}\right)\qquad (X_\Lambda \equiv \eta_\ls X^\Sigma)
\end{equation}
We then find, for the attractor condition
\begin{equation}
\partial_i V =0 \Rightarrow 2 \bar Z D_i Z =0
\end{equation}
which has two solutions.
Either
\begin{equation}
Z\neq 0 \quad D_i Z =0
\end{equation}
in which case the black-hole potential at the extremum is
\begin{equation}
V_{\mbox{extr}}=| Z|^2 = I_2
\end{equation}
or
\begin{equation}
Z = 0 \quad D_i Z \neq 0
\end{equation}
giving
\begin{equation}
V_{\mbox{extr}}=|D_i Z|^2 =- I_2.
\end{equation}
Here $I_2$ is the quadratic $U(1,n)$ invariant written in terms of the black-hole charge $(p_0,q_0,p_i,q_i)$, ($i=1,\cdots n$) as
\begin{equation}
I_2 = q_0^2 + p_0^2 - \sum_i (p_i^2 + q_i^2).
\end{equation}
From the analysis of \cite{bfgm}, the solution with $Z\neq 0$ exists
for $I_2 >0$ and  is a $N=2$, BPS critical point which is a genuine
attractor, since the Hessian matrix of the black-hole potential is
positive definite \cite{fk,fgk}. For the solution with $Z=0$,
which implies $I_2 <0$, the critical point is $N=2$ non-BPS and has
$2(n-1)$ flat directions since the Hessian matrix is semidefinite
positive with only two non vanishing eigenvalues. Note that for
$n=3$ the model can also be interpreted as the bosonic sector of
$N=3$ supergravity coupled to one vector multiplet \cite{maina}. In
the latter case, the BPS and non-BPS solutions are exchanged
\cite{veneziano} and the four flat directions of the BPS solution in
the $N=3$ model correspond to the hypermultiplet in the $N=3 \to
N=2$ decomposition.

For more general $N=2$ $\sigma$-models (with $C_{ijk}\neq 0$), to have a (anti-) holomorphic kinetic matrix a truncation in the matter sector is needed to satisfy eq. \eq{antiholn} such that  $\partial_i \cN_\ls=0$  \cite{adf01}.

\subsection{Genuine $N=1$ examples}
Among the class of $N=1$ supersymmetric theories with a
non-trivial electric-magnetic duality group, one can consider a
model with $n^2$ complex scalars coupled to $2n$ vector
multiplets. In this case the non-linear $\sigma$-model is the
K\"ahler manifold $U(n,n)/U(n)\times U(n)$ and the
electric-magnetic duality group is $U(n,n) \subset
Sp(4n,\mathbb{R})$, with the electric and magnetic field-strengths
embedded in the $2n+2\bar n$ of $U(n,n)$ \cite{az}. Denoting by
$s_{i\jbar}$ the holomorphic coordinates on the $\sigma$-model,
with $F^i_{\alpha\beta}= F^i_{\beta\alpha}$ the self-dual part (in
spinor notation) of the field strength of the complex vector $A^i$
and with $F^\ibar_{\alpha\beta}$ the self-dual part of the
field-strngth of the complex conjugate vector $A^\ibar =(A^i)^*$,
then the vector kinetic term is just
\begin{equation}
\cL =\Im \left(s_{i\jbar}F^i_{\alpha\beta} F^{\jbar\alpha\beta}\right).
\end{equation}
For $n=1$ this model coincides with the $N=2$ model previously considered and for $n=3$ it is the bosonic sector of $N=3$ supergravity coupled to three vector multiplets, while
for other $n$ it does not  have a higher $N$ origin.
As it was explicitly shown for $n=1$ and $n=3$, these models admit in general attractor black-hole solutions \cite{bfgm,veneziano}.


\section{$N=1$ examples as $N=2$ truncations}

The supersymmetry reduction of $N=2 \to N=1$ supergravity is obtained by truncating the $N=1$ spin 3/2 multiplet containing the second gravitino and the graviphoton.

All orientifold models in which the $N=1$ truncation leaves some  vectors and scalars with a non trivial holomorphic matrix $f_\ls =- \bar \cN_\ls$ (in the subspace which excludes the graviphoton) may be studied to see whether they have attractors or not.

A general analysis of the consistent truncation of $N=2$ theories to $N=1$
has been given in \cite{adf01}, to which we refer for all the details.
We just quote here the main results for the reduction of the vector multiplet sector.
Let us first  decompose the coordinates  of the $N=2$ special manifold as
$z^\cI \to (z^i, z^\alpha)$, with $i=1,\cdots n_S$  $N=1$ chiral multiplets while $\alpha$ labels the rest of the coordinates,
and the $N=2$ vectors as
$A^{\bf \Lambda}\to (A^\Lambda, A^X)$, with $\Lambda =1,\cdots n_V$ enumerating the $N=1$ vectors and $X$ the rest of the $N=2$ vectors.
 a consistent truncation requires, on the $N=1$ theory:
\begin{eqnarray}
&A^X=0 \,,\qquad z^\alpha = \mbox{const.}&\\
&L^\Lambda =0\,, \qquad f^\Lambda_i =0\,,\qquad f^X_\alpha=0&\\
&\cN_{\Lambda X}=0\,,\qquad C_{\alpha\beta\gamma}=0\,,\qquad C_{ij\alpha}=0&
\end{eqnarray}
which in particular imply the truncation of the graviphoton projector:
\begin{equation}
 T_{ \Lambda} = (\cN-\bar\cN)_{\Lambda\bf \Sigma}L^{\bf\Sigma} =0.
\end{equation}
This immediately shows that, whenever an $N=2$ holomorphic prepotential $F(X)$ exists such that $F_\Lambda = \partial F/\partial X^\Lambda$ (and $F_\ls \equiv \partial^2 F/\partial X^\Lambda \partial X^\Sigma$), the $N=1$ vector kinetic matrix is indeed anti-holomorphic, since in that case
\begin{equation}
\cN_\ls = \bar F_\ls -2\ii \bar T_\Lambda \bar T_\Sigma (L^{\bf\Gamma} \Im F_{\bf\gd}L^{\bf\Delta}) \to \bar F_\ls\,,\quad \partial_i \bar F_\ls=0
\end{equation}
so that we can identify $f_\ls $ with $- F_\ls$.
However, from the analysis of the previous section,  eq. \eq{antiholn},
it turns out that the matrix $\cN$ is always anti-holomorphic  in the reduced theory (even in the cases where no prepotentail $F$ exists) since $\Im \cN_{\Lambda\bf\Sigma}f^{\bf\Sigma}_i\to 0$.

\bigskip

An interesting possibility,  considered in \cite{adf01}, is the case of the $N=2$ theory based on the $\sigma$-model
\begin{equation}
\frac{SU(1,1)}{U(1)} \times \frac{SO(2,n)}{SO(2)\times SO(n)}.
\end{equation}

Let us study in detail the attractor equations for the $N=1$ truncation of this model where only the dilaton chiral multiplet is kept together with $n$ vector multiplets. In this case the kinetic matrix simply becomes
\begin{equation}
f_\ls = S \delta_\ls
\end{equation}
and the duality group reduces to $SL(2,\mathbb{R}))\times SO(n)\subset
Sp(2n,\mathbb{R})$, where $SL(2,\mathbb{R}))$ acts as electric-magnetic
duality.

Referring to the discussion in section \ref{embedding}, we have in this case
\begin{equation} \cV_\Lambda = q_\Lambda -S p_\Lambda
\end{equation} and, from \eq{min}
\begin{equation}
\partial_S V =0 \Rightarrow \sum_\Lambda \cV_\Lambda \cV_\Lambda =0. \label{attr0}
\end{equation}
An attractor solution is then found for
\begin{equation}
S =a+\ii b\,:\qquad a=-p\cdot q /p^2 \,;\quad b=-\sqrt{p^2q^2 -(p\cdot q)^2}/p^2.
\end{equation}
and, substituting in the extremized black-hole potential gives
\begin{equation}
V|_{\mbox{extr}}=  \sqrt{p^2q^2 -(p\cdot q)^2}, \label{entro3}
\end{equation}
with a positive Hessian matrix, since:
\begin{equation}
\frac{\partial^2 V}{\partial a^2}|_{\mbox{extr}} = \frac{\partial^2 V}{\partial b^2}|_{\mbox{extr}} = \frac {(p^2)^2}{\sqrt{p^2q^2 -(p\cdot q)^2}}\,; \qquad \frac{\partial^2 V}{\partial a \partial b}|_{\mbox{extr}} =0.
\end{equation}

Note that the entropy for this model has formally the same expression (with $SO(6,n)$ replacing $SO(n)$)
 as in the general $N=4$ theory \cite{cvy,dlr,s,fk} since the non trivial electric-magnetic duality $SL(2,\mathbb{R}))$ is the same.

For $n=6$, the bosonic sector of this model coincides with the bosonic sector  of pure $N=4$ supergravity.

For $n=2$,  its bosonic sector coincides instead with the one of the $N=2$ theory of the quadratic series, with one ($N=2$) vector multiplet.
Note in fact that the quartic invariant $I_4=q^2p^2 -(q\cdot p)^2$ reduces in this case (where we only have $q_0,p^0,q_1,p^1$) to the square of the quadratic invariant $I_2 =p^0q_1-p^1q_0$, $I_4 =(I_2)^2$.

For $n=1$, the quartic invariant is zero, since in this case
$(q\cdot p)^2 = q^2 p^2$. This case concides with the first
example of Section 2 ( the $Sp(2n,\mathbb{R})/U(n)$ series) for
$n=1$.

 \subsection{CY orientifold compactifications and $N=1$ reduction of  homogeneous $N=2$  models}

The model discussed above may be generalized by considering the compactification to four dimensions  of Type IIA theory on orientifolds  (or of M-theory on a special class of $G_2$-manifolds), as discussed in \cite{gl}.
According to \cite{gl}, by considering a Type IIA orientifold which keeps only the
complex K\"ahler moduli $z^A$ and vectors $A^\alpha_\mu$, with
$A=1,\dots , h_{1,1}^-$ and $\alpha =1,\dots , h_{1,1}^+$, the $N=1$ kinetic
matrix  for the bulk vectors has the simple
form (which generalizes the expression for the 1-modulus $S$ case)
\begin{eqnarray}
f_{\alpha\beta} =-\bar\mathcal{N}_{\alpha\beta}&=&{z}^A\,d_{A{\alpha\beta}}\,. \label{orient}
\end{eqnarray}
Similar expressions exist also for the gauge kinetic matrix of the brane vectors (as a function of the bulk moduli) \cite{gl,lust}
So one could consider the example of a truncation of the homogeneous (but non symmetric) space
$L(0,P,\dot{P})$  in which $z^A=(S,z^2, z^3)$ and
$z^\alpha=(z^m,z^{\dot{m}})$ ($m=1,\dots, P$ and $\dot{m}=1,\dots,
\dot{P} $). In this theory the only non vanishing
entries for the $d$ tensor are
\begin{eqnarray}
d_{S22}&=&-d_{S33}=\frac{1}{2}\,\,\,;\,\,\,\,d_{2mn}=d_{3mn}=\delta_{mn}\,\,\,;\,\,\,\,
d_{2\dot{m}\dot{n}}=-d_{3\dot{m}\dot{n}}=\delta_{\dot{m}\dot{n}}\,.\nonumber\\&&
\end{eqnarray}
In the orientifolded theory we would have
$z^A=(S,z^2,z^3)$ and $A^\alpha_\mu=(A^m_\mu, A^{\dot{m}}_\mu)$.

Let us now analyze the attractor behavior of the $N=1$ reduction for more general  homogeneous $L(q,P,\dot P)$ Special K\"ahler models \cite{dwvsvp}.
These models have $r+q+3$ complex scalars, with $r=(P+\dot P)\mathcal{D}_{q+1}$,
$\mathcal{D}_{q+1}$ being the irreducible reprsentation of $spin(1,q+1)$.
The truncation to $N=1$ leaves $q+2$ chiral multiplets together with $r$ vector multiplets.
In particular, the scalar $S$ corresponding to the dilaton decouples from the rest and after the orientifold
projection we are left with the coordinates $z^A=x^A+i\,y^A$, $A=0,1,\dots, q+1$ spanning the $\sigma$-model $SO(2,q+2) /[SO(2)\times SO(q+2)]$.
Let us denote the vector fields by $A^\alpha$, $\alpha=1,\dots, r$.

The  holomorphic kinetic matrix is now a particular case of \eq{orient}; written in terms of the $\gamma$-matrices of $SO(1,q+1)$, it is:
\begin{eqnarray}
f_{\alpha\beta}&=&{z}^A\,{\bf \Gamma}_{A\alpha\beta}\,,
\end{eqnarray}
where
\begin{eqnarray}
{\bf \Gamma}_{0\alpha\beta}&=&-\delta_{\alpha\beta}\,\,;\,\,\,\,{\bf \Gamma}_{i
\alpha\beta}=(\gamma_i)_{\alpha\beta}\,,\quad i=1,\dots q+1\nonumber\\
\overline{{\bf
\Gamma}}_{0}^{\alpha\beta}&=&\delta^{\alpha\beta}\,\,;\,\,\,\,\overline{\bf
\Gamma}_{i}^{
\alpha\beta}=(\gamma_i)_{\alpha\beta}\,,\nonumber\\
{\bf \Gamma}_{(A}\,\overline{{\bf
\Gamma}}_{B)}&=&\eta_{AB}\,\,\,;\,\,\,\,\eta={\rm
diag}(-1,+1,\dots,+1)\,.\label{cliff}
\end{eqnarray}
are two copies of $SO(1,q+1)$ $\gamma$-matrices.
They together  compose the $2r\times 2r$ representation of the ${\rm SO}(1,q+1)$ gamma
matrices, corresponding to the embedding in the electric-magnetic duality group $SO(2,q+2)$, which reads
\begin{eqnarray}
\Gamma_A&=&\left(\matrix{{\bf 0} & {\bf \Gamma}_{A}\cr
\overline{\bf \Gamma}_A & {\bf 0}}\right)\, ,
\end{eqnarray}
The above equations are in fact written for the case $P=1$, $\dot P=0$. An obvious extension is understood for $P$, $\dot P$ generic (when this is the case, in \eq{cliff}  $P \to \dot P$ requires ${\bf \Gamma}_i \to - {\bf
\Gamma}_i $) and will be used in section \ref{L0}.

So $\mathfrak{so}(1,q+1)$ is an electric subalgebra of the
electric-magnetic algebra $\mathfrak{so}(2,q+2)$, and the system
of electric and magnetic field-strengths $
\cS=(F^\alpha,G_\alpha)$ compose the spinor representation of
$SO(2,q+2)$. To be more precise,  the (real) spinor of electric
and magnetic charges is irreducible under $SO(2,q+2)$ but
decomposes as $\cS = \cS^+_e + \cS^-_m$ for $SO(2,q+2) \to
SO(1,q+1)\times SO(1,1)$, where $\cS^\pm$ have opposite grading
under $SO(1,1)$ and, for $q$ even, also opposite chirality. The
$2r$-dimensional $SO(2,q+2)$ spinorial representation can be
described in terms of the following $2r\times 2r $ matrices
$\Gamma_M=\{\Gamma_{-1},\,\Gamma_A,\,\Gamma_{q+2}\}$ and
$\overline{\Gamma}_M=\{-\Gamma_{-1},\,\Gamma_A,\,\Gamma_{q+2}\}$
($M,\,N=-1,\dots,\,q+2$), where
\begin{eqnarray}
\Gamma_{-1}&=&\bfone_r\times
\bfone_2\,\,;\,\,\,\Gamma_{q+2}=\bfone_r\times \sigma_3\,,
\end{eqnarray}
which satisfy the relations
\begin{eqnarray}
\Gamma_{(M}\overline{\Gamma}_{N)}=\hat{\eta}_{MN}\,\,\,;\,\,\,\,\hat{\eta}={\rm
diag}(-1,-1,+1,\dots,+1)\,.
\end{eqnarray}
The action of the $\mathfrak{so}(2,q+2)$ generators on the $2r$
electric-magnetic charges is defined by the matrices
$J_{MN}=\frac{1}{4}\,\Gamma_{[M}\overline{\Gamma}_{N]}$.\par
 The K\"ahler potential in a special-coordinate
inspired basis is
\begin{equation}
K=-\log Y
\end{equation}
with
\begin{eqnarray}
Y &=& -\frac{1}4 \left[(z_0-\bar z_0)^2 - (z_i -\bar z_i)^2 \right]\,,\quad i=1,\cdots q+1\nn\\
&=&  -\frac{1}4 \eta_{AB} (z^A -\bar z^A)(z^B -\bar z^B) \equiv ||y||^2
\end{eqnarray}
and, in terms of $z^A=x^A+\ii y^A$, the vector kinetic matrix is
\begin{equation}
f_{\alpha\beta} = ({\bf \Gamma} _A)_{\alpha\beta} z^A \,, \qquad \Im f_{\alpha\beta}<0.
\end{equation}

 We find
also:
\begin{eqnarray}
{\rm Im} f_{\alpha\beta}&=& y^A\,{\bf
\Gamma}_{A\alpha\beta}\,\,\,;\,\,\,\,\,{\rm Im}
f^{-1\,\alpha\beta}=\frac{y^A}{||y||^2}\,\overline{{\bf
\Gamma}}_{A}^{\alpha\beta}\,\,\,;\,\,\,\,\,||y||^2=y^A\eta_{AB}y^B
<0\,,\nonumber\\&&
\end{eqnarray}
The black-hole potential reads \footnote{For the $L(q,P,\dot P)$ models with $P\dot P \neq  0$ ($q=4m$), since $\Im f_{\alpha\beta}$ is block-diagonal in the $P,\dot P$ space, two terms in two separate spinor spaces are understood in eq. \eq{potlqp}.}:
\begin{eqnarray}
V&=&-\frac 12 (q_\alpha - {z}^A{\bf \Gamma}_{A\alpha\gamma}p^\gamma){\rm Im}
f^{-1\,\alpha\beta}(q_\beta -\bar{z}^A{\bf
\Gamma}_{A\beta\delta}p^\delta)=\nonumber\\
&=&-\frac{1}{2||y||^2}\left(y\cdot N-2\,x^T\,W\,y+(y\cdot
M)(||y||^2-||x||^2)+2\,(x\cdot M)(x\cdot
y)\right)\,,\nonumber\\&&\label{potlqp}
\end{eqnarray}
where we have introduced the following shorthand notation
\begin{eqnarray}
N_A&=&q_\alpha \overline{{\bf
\Gamma}}_{A}^{\alpha\beta}q_\beta \,\,;\,\,\,\,M_A=p^\alpha{\bf
\Gamma}_{A\alpha\beta}p^\beta \,\,\,;\,\,\,\,W_{AB}=p^\alpha ({\bf
\Gamma}_{A}\overline{{\bf \Gamma}}_B)_\alpha{}^\beta q_\beta\,,\nonumber\\
y\cdot M&\equiv & y^A M_A\,\,\,;\,\,\,\,x\cdot
y=x^A\,\eta_{AB}\,y^B\,\,\,;\,\,\,\,x^T\,W\,y=x^A\,W_{AB}\,y^B\,.
\end{eqnarray}
The extremization condition may be written in the elegant form
\begin{equation}
\bar\cV^\alpha {\bf \Gamma}_{A\alpha\beta}\bar\cV^\beta =0
,\label{purespinor}
\end{equation}
in terms of the spinor
\begin{equation}
\bar\cV^\alpha = \frac{1}{||y||^2}\left(y^A\, q_\beta
(\overline{{\bf \Gamma}}_A)^{\beta\alpha}-x^A\,y^B\,p^\beta\,({\bf
\Gamma}_A\overline{{\bf \Gamma}}_B)_\beta{}^\alpha
\right)+\ii\,p^\alpha.\label{spinor}
\end{equation}
Equation (\ref{purespinor}) can be written as the following real
conditions in the real and imaginary parts of the $z^A$ moduli:
\begin{eqnarray}
&&N_A+M_A\,(||y||^2-|x||^2)-2\,y_A\,(y\cdot
M)+2\,W_{[AD]}\,x^D=0\,,\nonumber\\
&&x_A=\frac{(p\cdot q)}{||M||^2}\,M_A-\frac{1}{(y\cdot
M)}\,P_A{}^B\,W_{[BC]}\,y^C\,,
\end{eqnarray}
where $P_A{}^B$ is the projector in the directions orthogonal to
$M_A$:
\begin{eqnarray}
P_A{}^B&=&\delta_A^B-\frac{M_A\,M^B}{||M||^2}\,.
\end{eqnarray}
 Eq.s \eq{inverseformula}, \eq{spinor} and \eq{purespinor}
allow to write down a general expression for the entropy, given by
eq. \eq{entropy}:
\begin{equation}
\frac 1\pi S_{BH}(p,q)= V|_{\partial_i V =0} = -\, M_A\,
y^A|_{extr}\,. \label{entropylpq}
\end{equation}

Geometrically, the attractor points are the points where
$\bar\cV^\alpha$ becomes a pure spinor.\footnote{Here we adopt a
definition \cite{purespin1,purespin2,dft} which is milder than the
mathematical definition when $q>8$ (and $P,\,\dot P>1$)} As a
remark we observe that eq. (\ref{purespinor}) is identical in form
to eq. (4.43) of \cite{dft} for the $N=2$ attractors of
homogeneous K\"ahler spaces with vanishing central charge (and
vanishing of the $q+2$ matter charges $Z_I$). On a general ground
this is a consequence of the fact that the $N=1$ attractor
equations given in eq. (\ref{min}) are similar to the ones for the
$N=2$ attractors (eq. \eq{special}) with vanishing central change:
\begin{eqnarray}
C_{ijk}\,\overline{Z}^{\,\,j}\overline{Z}^{\,\,k}&=&0\,,
\end{eqnarray}
if one replaces $C_{ijk}$ by $\partial_i f_{\alpha\beta}$ and
$\overline{Z}^{\,\,i}=g^{i\bar\jmath}D_{\bar\jmath}\,\overline{Z}$
by $\overline{\mathcal{V}}^{\,\alpha}={\rm Im}
f^{-1\,\alpha\beta}\,\overline{\mathcal{V}}_\beta$.\par As the
above discussion shows, $L(q,P,\dot P)$ theories may admit in
general attractor extrema, apart from particular cases. We are
going to discuss, in the rest of this section, some specific
examples.

\subsubsection{The $L(q,1)$ cases.}

This series, for particular values of $q$: $q=1,2,4,8$, describes $N=2$ symmetric spaces \cite{gst}. Let us consider in particular the case $q=8$, which corresponds to the $\sigma$-model $E_{7(-25)}/E_6\times U(1)$, when decomposed with respect to $SL(2,\mathbb{R}))\times SO(2,10)$ in a truncation where one only keeps the $SL(2,\mathbb{R}))$ singlets. Since the representation of the electric and magnetic field-strengths decomposes under $SL(2,\mathbb{R}))\times SO(2,10)$ as
\begin{equation}
56\to (2,12)+(1,32),
\end{equation}
only the 32 electric and magnetic field-strengths belonging to the spinorial representation of $SO(2,10)$ are kept.
In the $\sigma$-model counterpart
\begin{equation}
E_{7(-25)}/[E_6\times U(1)] \to SO(2,10)/[SO(2)\times SO(10)].
\end{equation}
So, also in this case the final $N=1$ model is based on the $SO(2,10)/[SO(2)\times SO(10)]$ $\sigma$-model coupled to $F,G$ in the spinorial representation of $SO(2,n+2)$, with electric subalgebra $SO(1,n+1)$.

For the $L(2,1)$ model, the gamma matrices read
\begin{eqnarray}
{\bf \Gamma}_1&=&-\left(\matrix{ 1 & 0 & 0 & 0 \cr 0 & 1 & 0 & 0
\cr 0 & 0 & 1 & 0 \cr 0 & 0 & 0 & 1 \cr  }\right)\,\,;\,\,\,{\bf
\Gamma}_2=\left(\matrix{ 0 & 0 & 0 & 1 \cr 0 & 0 & 1 & 0 \cr 0 & 1 & 0 & 0 \cr 1 & 0 & 0 & 0 \cr  }\right)\,,
\nonumber\\
{\bf \Gamma}_3&=&\left(\matrix{ 0 & 0 & 1 & 0 \cr 0 & 0 & 0 & -1
\cr 1 & 0 & 0 & 0 \cr 0 & -1 & 0 & 0 \cr
}\right)\,\,\,;\,\,\,\,{\bf \Gamma}_4=\left(\matrix{ 1 & 0 & 0 &
0 \cr 0 & 1 & 0 & 0 \cr 0 & 0 & -1 & 0 \cr 0 & 0 & 0 & -1 \cr
}\right)\,.
\end{eqnarray}
The potential at the extremum has the following expression:
\begin{eqnarray}
V_{extr}&=&(p^1\, q_2 -p^2\,q_1+ p^4\,q_3-p^3\,q_4)\,.
\end{eqnarray}
 This agrees with the fact that the charge-spinor in this case belongs to the $\mathbf{4} \in SU(2,2) = spin(4,2)$, which is complex, while the entropy is given in terms of a real bilinear invariant \cite{hel}.

The $L(1,1)$ case has no attractors, as we will see in the following, as a particular case of the series $L(1,P)$.

Let us now move to analyze various cases of homogeneous spaces $L(q,P)$ and $L(q,P,\dot P)$.

\subsubsection{The $L(0,P,\dot P)$ cases.} \label{L0}
For this series ($q=0$), the spinor of charges degenerates, and we have $P+ \dot P$
spinorial electric and magnetic charges $(p^\alpha ,\,q_\alpha)$ and $(\dot p^\alpha ,\,\dot q_\alpha)$. The gamma
matrices read
\begin{eqnarray}
&{\bf \Gamma}_0 =\pmatrix{-\delta_{\alpha\beta} &0\cr 0&
-\delta_{\dot\alpha\dot\beta} }= -\bar{\bf \Gamma}_0\;;\qquad {\bf
\Gamma}_1 =\pmatrix{\delta_{\alpha\beta} &0\cr 0&
-\delta_{\dot\alpha\dot\beta} }= \bar{\bf \Gamma}_1\,,&
\end{eqnarray}
so that the scalar potential is
\begin{eqnarray}
V&=&  -\frac 12 \cV^\alpha \Im f_{\alpha\beta} \bar \cV^\beta
-\frac 12 \cV^{\dot\alpha} \Im f_{\dot\alpha\dot\beta}
\bar\cV^{\dot\beta}\label{potlopp}
\end{eqnarray}
with
\begin{eqnarray}
\Im f_{\alpha\beta} =-\delta_{\alpha\beta} (y_0-y_1) &;& \cV^\alpha = -\frac 1{y_0-y_1}\left[q^\alpha +(z_0-z_1) p^\alpha\right]; \\
\Im f_{\dot\alpha\dot\beta} =-\delta_{\dot\alpha\dot\beta} (y_0+y_1) &;& \cV^{\dot\alpha} = -\frac 1{y_0+y_1}\left[q^{\dot\alpha} +(z_0+z_1) p^{\dot\alpha}\right].
\end{eqnarray}
To have $\Im f <0$ requires $y_0>y_1 >0$.

For this series, the potential \eq{potlopp} decomposes into  the sum of two independent, functionally identical, contributions, each one depending on a different variable:
\begin{equation}
V=  V(u) + \dot V(\dot u) \,; \qquad u\equiv z_0-z_1\,, \quad \dot u=z_0+z_1 \label{pot2lopp}
\end{equation}
where:
\begin{eqnarray}
 V(u)&=&-\frac 12 \cV^\alpha \Im f_{\alpha\beta}\bar\cV^\beta (u)= \frac 1{2\,{\rm Im} u} \left( q^2 + 2\, (q\cdot p) \,\Re\, u+  p^2 \,|u|^2\right)\nonumber\\
\dot V({\dot u})&=&
-\frac 12 \cV^{\dot\alpha} \Im f_{\dot\alpha\dot\beta}
 \bar\cV^{\dot\beta}({\dot u})
= \frac 1{2\,{\rm Im} \dot u} \left(\dot  q^2 + 2\, (\dot q\cdot
\dot p) \,\Re\, \dot u+ \dot  p^2 \,|\dot u|^2\right)
\end{eqnarray}
The attractor equations become the  equations for two cones, which
can be regarded as the pure spinor equations for $SO(1,1)$:
\begin{eqnarray}
\sum_{\alpha =1,\cdots P} \cV^\alpha \cV^\alpha &=&0 \\
\sum_{\dot\alpha =1,\cdots \dot P} \cV^{\dot\alpha} \cV^{\dot\alpha} &=&0.
\end{eqnarray}
Therefore the attractor points are the ones for which the complex
vectors $\cV^\alpha$ and $\cV^{\dot\alpha}$ have vanishing euclidean
norm. The minima of \eq{pot2lopp} are found for
\begin{eqnarray}
u&=&- \frac 1{p^2} \left( q\cdot p -\ii\sqrt{I_4}\right)\\
\dot u&=& -\frac 1{\dot p^2} \left( \dot q\cdot \dot p
-\ii\sqrt{\dot I_4}\right)
\end{eqnarray}
where $I_4 \equiv q^2p^2-(q\cdot p)^2$, $\dot I_4 \equiv \dot q^2\dot p^2-(\dot q\cdot \dot p)^2$.
The extremum of the black-hole potential is then
\begin{eqnarray}
V|_{\mbox{extr}}= \sqrt{I_4}+\sqrt{\dot I_4}.
\end{eqnarray}
The Hessian matrix at the extremum, evaluated with respect to the real and imaginary parts of $u,\dot u$, is
\begin{eqnarray}
H(u,\dot u)|_{\mbox{extr}} =\pmatrix{ \frac{p^4}{\sqrt{I_4}}\pmatrix{1&0\cr 0&1} &0\cr
0&\frac{\dot p^4}{\sqrt{\dot I_4}}\pmatrix{1&0\cr 0&1} }\,,\qquad \det H|_{\mbox{extr}}>0.
\end{eqnarray}
showing that for all this class of models the extrema of the potential have indeed an attractor nature.

Note that for all $L(0,P,\dot P)$ the duality group is $SO(2,2)\times SO(P)\times SO(\dot P)$, and the potential at the extremum  may be written in terms of the manifest invariant of the duality group
 \begin{equation}
I_4=T_{\alpha\beta}T^{\alpha\beta} = p^2 q^2 -(p\cdot q)^2, \label{0pp}
 \end{equation}
(and similarly for $\dot I_4$),  with $T_{\alpha\beta} =-T_{\beta\alpha}\equiv \cS_{\alpha}^T \cdot \Omega \cdot \cS_{\beta}$,  $\cS^T_\alpha =(p^\alpha,q_\alpha)$  an $SO(P)$-valued  chiral spinor of $SO(2,2)$ and $\Omega$ the invariant metric of $SU(1,1)\subset SO(2,2)$.
This class of models is particularly interesting because it may correspond to a system of $P$ $D3$ and $\dot P$ $D7$ branes on Calabi--Yau orientifold compactifications \cite{adft}.

If $P\dot P \neq 0$, both $P$ and $\dot P$ must be bigger than one, otherwise the attractor point does not exist (since then $I_4$ or $\dot I_4$ vanish, and $\Im\,u$ or $\Im \,\dot u$ would vanish either.).

For $P\dot P=0$, we have the $L(0,P)$ ($P>1$) models, in which case one complex  modulus ($u$ or $\dot u$) is undetermined on the black-hole solution, the Hessian has two vanishing eigenvalues and the attractor equations have two flat directions.

Let us finally observe that, since the irreducible representation of the spinor of charges in $SO(2,2)$ is in fact  chiral, only a subgroup $SL(2,\mathbb{R}))\times SO(P)\times SO(\dot P)\subset SO(2,2)\times SO(P)\times SO(\dot P)$ of the duality group acts non trivially. The vector-multiplet sector of this theory (in the case $\dot P=0$) is then identical to the $N=1$ truncation of the $L(-1,P)$ series. In this last case, however, the scalar sector reduces to the coset $SU(1,1)/U(1)$, so that the attractor condition is one complex equation for one modulus.
 Then the critical point is a genuine attractor.
Note that this truncation gives back the same $SU(1,1)/U(1)\times SO(P)$ model already discussed in section 3, whose entropy has been given in \eq{entro3}.

\subsubsection{The $L(1,2)$ case.} We have four electric and four
magnetic charges. The gamma matrices read:
\begin{eqnarray}
\hskip -2mm{\bf \Gamma}_1&=&-\left(\matrix{ 1 & 0 & 0 & 0 \cr 0 & 1 & 0 & 0
\cr 0 & 0 & 1 & 0 \cr 0 & 0 & 0 & 1 \cr  }\right);\,{\bf
\Gamma}_2=\left(\matrix{ 1 & 0 & 0 & 0 \cr 0 & -1 & 0 & 0 \cr 0 &
0 & 1 & 0 \cr 0 & 0 & 0 & -1 \cr  }\right);\,{\bf
\Gamma}_3=\left(\matrix{ 0 & 1 & 0 & 0 \cr 1 & 0 & 0 & 0 \cr 0 &
0 & 0 & 1 \cr 0 & 0 & 1 & 0 \cr  }\right)\,.\nonumber\\
\end{eqnarray}
The potential at the extremum reads
\begin{eqnarray}
V_{extr}&=&(p^2\,q_4-q_2\,p^4+p^1\,q_3 -q_1\,p^3)\,.
\end{eqnarray}
From this we see that in the symmetric $L(1,1)$ case, in which
$p^3=q_3=p^4=q_4=0$, the potential at the extremum is zero. This
is in agreement with the fact that the in this case we have a
single spinor of charges, belonging to the $\mathbf{4} \in
Sp(4,\mathbb{R}) = spin(3,2)$, which has no antisymmetric bilinear
invariant.

\section{Concluding remarks} \label{conclusions}
In this investigation we have considered the black-hole potential
of charged extremal black-holes in $N=1$ supergravity coupled to
chiral and Maxwell vector multiplets. The attractor equations take
the particular simple form \eq{min}. In a particular class of
models, obtained by an orientifold projection of homogeneous
special geometries, the attractor equation (\ref{purespinor}) has
the geometrical meaning, at least for $q \leq 8$,  that the spinor
$\cV^\alpha$ defined in \eq{spinor} is a pure spinor. Pure spinors
have already occurred in the literature in connection to attractor
equations for type II compactifications on generalized Calabi--Yau
manifolds in \cite{tom}.

The entropy can be computed and it is given
in terms of invariants of the electric-magnetic duality group
that, for an $N=1$ reduction of $L(q,P,\dot P)$ homogeneous
spaces, is in general $spin(2,q+2)\times S_q(P,\dot P)$, where
$S_q(P,\dot P)$ is the centralizer of the relevant Clifford
algebra and it was classified in \cite{dwvsvp}.
 For models of the type $L(0,P,\dot P)$, the underlying
special geometry may correspond to D-branes
 on a CY-orientifold compactification and the attractor
 points would correspond to extremal black-holes on the branes.
From the analysis of section 3, we find that such attractors exist
if at least two branes of the same kind are kept. We also find
evidence that extremal black-holes with attractor behavior may
exist in heterotic string compactifications on Calabi--Yau
manifolds, with the dilaton and axion fields fixed in terms of the
electric and magnetic charges of the vector bundle. This is the
$N=1$ analogue of the $N=4$ dilaton-axion black-hole
\cite{kp,kop,bko}. On a more general ground, it seems that,
whenever the gauge-kinetic matrix is moduli-dependent in $N=1$
supergravity models coupled to vector multiplets, then charged
extremal black-hole solutions with attractor behavior appear as a
generic rather than an exceptional feature.

It would be interesting to extend the present analysis by including deviations from the Maxwell--Einstein system, by considering either Born--Infeld contributions to the Maxwell action \cite{gr,cst,az,bi}  (as it would be relevant in the case of brane vector fields) and higher curvature terms \cite{wald,moh,sen,dew,pio} in the gravitational field.


\section*{Acknowledgements}
We acknowledge enlightening discussions with C. Angelantonj, I. Antoniadis, P.
Aschieri and R. Stora.
 Work supported in part by the European Community's Human Potential
Program under contract MRTN-CT-2004-005104 `Constituents,
fundamental forces and symmetries of the universe', in which L.A.,
R.D'A. and M.T. are associated to Torino University. The work of
S.F. has been supported in part by European Community's Human
Potential Program under contract MRTN-CT-2004-005104 `Constituents,
fundamental forces and symmetries of the universe' and the contract
MRTN- CT-2004-503369 `The quest for unification: Theory Confronts
Experiments', in association with INFN Frascati National
Laboratories and by D.O.E. grant DE-FG03-91ER40662, Task C.

\end{document}